\documentclass[aps,prapplied,twocolumn,byrevtex,showpacs,floatfix,superscriptaddress,reprint]{revtex4-2}
\usepackage[utf8]{inputenc}
\usepackage[english]{babel}
\usepackage[T1]{fontenc}
\usepackage{amsmath, amssymb, amsfonts}
\usepackage{tikz}
\usepackage{graphicx}
\usepackage{color}
\usepackage[colorlinks=true,linkcolor=blue,citecolor=magenta,urlcolor=blue]{hyperref}
\usepackage{silence,lmodern}
\usepackage{caption}
\usepackage{support-caption}
\usepackage{subcaption}
\usepackage{physics}

\begin{document}


\title{Variational determination of multi-qubit geometrical entanglement in NISQ computers}



\author{A.~D.~Muñoz-Moller}
\email{andmunoz2016@udec.cl}
\affiliation{Instituto Milenio de Investigaci\'on en \'Optica y Departamento de F\'isica, Facultad de Ciencias F\'isicas y Matem\'aticas, Universidad de Concepci\'on, Casilla 160-C, Concepci\'on, Chile}
\author{L.~Pereira}
\affiliation{Instituto de F\'{\i}sica Fundamental IFF-CSIC, Calle Serrano 113b, Madrid 28006, Spain}
\author{L.~Zambrano}
\affiliation{Instituto Milenio de Investigaci\'on en \'Optica y Departamento de F\'isica, Facultad de Ciencias F\'isicas y Matem\'aticas, Universidad de Concepci\'on, Casilla 160-C, Concepci\'on, Chile}
\author{J.~Cort\'es-Vega}
\affiliation{Instituto Milenio de Investigaci\'on en \'Optica y Departamento de F\'isica, Facultad de Ciencias F\'isicas y Matem\'aticas, Universidad de Concepci\'on, Casilla 160-C, Concepci\'on, Chile}
\author{A.~Delgado}
\affiliation{Instituto Milenio de Investigaci\'on en \'Optica y Departamento de F\'isica, Facultad de Ciencias F\'isicas y Matem\'aticas, Universidad de Concepci\'on, Casilla 160-C, Concepci\'on, Chile}

\begin{abstract}
Current noise levels in physical realizations of qubits and quantum operations limit the applicability of conventional methods to characterize entanglement. In this adverse scenario, we follow a quantum variational approach to estimate the geometric measure of entanglement of multiqubit pure states. The algorithm requires only single-qubit gates and measurements, so it is well suited for NISQ devices. This is demonstrated by successfully implementing the method on IBM Quantum devices for Greenberger-Horne-Zeilinger states of $3$, $4$, and $5$ qubits. Numerical simulations with random states show the robustness and accuracy of the method. The scalability of the protocol is numerically demonstrated via matrix product states techniques up to $25$ qubits.
\end{abstract}
\maketitle
\section{Introduction}
Entangled quantum states \cite{Schrod35}, which were first discussed as an argument against the completeness of quantum mechanics \cite{EPR}, are considered nowadays a distinguishing feature of this theory. Entanglement is also widely recognized as one of the core resources in the study of quantum computing and quantum information \cite{Bennett95}; the success of quantum algorithms such as Shor's algorithm \cite{Shors} is tied to proper implementations of non-local gates in quantum computers, while protocols like quantum teleportation \cite{QTele} and quantum key distribution \cite{QKD} rely on entangled states between two or more parties. Entanglement is also needed for both quantum sensing \cite{sensing} and quantum metrology \cite{metro1, metro2}.

Due to the role played by entanglement in the understanding of quantum mechanics and its many feasible applications, several methods have been developed to quantify and identify it. The Peres-Horodecki theorem \cite{Peres96, Horodecki96}, which is based on the negativity of the partial transpose mapping, allows to prove the existence of entanglement in qubit-qubit and qubit-qutrit pure or mixed quantum states, but it is inconclusive for higher dimensional bipartite systems. Along with this, partial transpose mappings correspond to nonphysical operations and therefore cannot be directly implemented experimentally. The Bell inequality \cite{Bell} can also be employed to detect entanglement of known states, which requires solving an optimization problem. Here, a violation signals the presence of entanglement. However, as the example of Werner states shows \cite{werner}, there are entangled states that do not violate Bell's inequality. Entanglement witnesses \cite{Horodecki96, Terhal2000} detect entanglement, although only for restricted states and without providing a meaningful answer for arbitrary states. Several experimental implementations have been proposed \cite{wit1, wit2, wit3}, but in order to construct a suitable witness some \textit{a priori} knowledge of the state is required. Additionally, an entanglement witness is based on the measurement of an observable that contain entangled states in its spectral decomposition, which makes experimental realizations difficult. Relative entropy of entanglement \cite{re1} has been shown to satisfy the monotonicity condition and therefore it is thought to be a good measure of entanglement. Nevertheless, it is very difficult to compute \cite{re2} and it is known only for a handful of examples, mainly bipartite pure states \cite{re3,re4}. Similarly, concurrence \cite{conc1} was first introduced as a measurement of entanglement for two-qubit mixed states, with the advantage of having a computable formula for entanglement of formation, which can be implemented experimentally \cite{conc2, conc3, conc4}. 

Here, we address the problem of experimentally measuring the amount of entanglement of multiqubit pure states in a noisy intermediate-scale quantum (NISQ) \cite{NISQ} computer. We use the geometric measure of entanglement (GME), an interesting entanglement monotone first introduced by Shimony \cite{gme1} for bipartite pure states and later generalized by Barnum and Linden \cite{gme2} to the multipartite case. This entanglement measure characterizes the entanglement of a pure state $|\psi\rangle$ as the distance to the nearest separable pure state $|\phi\rangle$ and it has been related to the relative entropy of entanglement \cite{wei_rent} and to entanglement witness \cite{WG2003}. The GME has been used to quantify the difficulty to distinguish multipartite quantum states using local operations \cite{Hayashi2006}, to study quantum phase transitions in spin models \cite{Ors2008} and to quantify how well a state serves as the input to Grover’s search algorithm \cite{Biham2002}. 

The GME of a state $|\psi\rangle$ can be obtained by maximizing the fidelity $F(|\psi\rangle,|\phi\rangle)=|\langle\psi|\phi\rangle|^2$ in the set $\{|\phi\rangle\}$ of separable pure states. This is an experimentally accessible quantity, since its evaluation only requires local measurements. In particular, the fidelity can be efficiently measured in a NISQ computer. This class of devices is characterized by a low number of available qubits, limited connectivity between qubits, low coherence times, and noisy entangling gates that restrict the depth of the circuits. Within this adverse scenario, variational quantum algorithms (VQA) are among the most popular \cite{vqa1, vqa2} strategies to achieve quantum advantage, showing promising results in quantum chemistry to find large Hamiltonian eigenvalues \cite{vqe1, vqe2, vqe3, vqe4}, solving tasks of quantum metrology \cite{varmetro1, varmetro2} and being applied in quantum machine learning \cite{qml}. 

In this article, we propose a variational determination of Geometrical Entanglement (VDGE). The algorithm employs a quantum device to estimate the fidelity, which is used to optimize a parameterized quantum circuit with a classical optimization method. This circuit iteratively implements projections of the state $\ket{\psi}$ on a sequence of multiqubit separable states that are generated by an optimization algorithm. We use the complex simultaneous perturbation stochastic approximation (CSPSA) algorithm as classical optimizer, which was first introduced in the context of state estimation \cite{URND2019, cspsa2}. This method exhibits mean convergence and is robust against noise. We show that our method correctly reproduces the value of the GME for arbitrary superpositions of Greenberger-Horne-Zeilinger (GHZ) and W states. Using Monte Carlo numerical experiments we study the overall accuracy achieved by our method as a function of the number of iterations for states of $n=2,3,4,5,6$ qubits. The feasibility of our approach to characterize the entanglement of states of larger numbers of qubits is also studied by means of matrix product state (MPS) techniques. Finally, we present the experimental results obtained by applying our method to the measurement of the GME of a GHZ state in IBM Quantum Falcon Processors \cite{IBMQ} \texttt{ibmq\_lima} and \texttt{ibmq\_bogota} for $n=3,4,5$ qubits.

\section{Theoretical background}

Let us consider a $n$-qubit system described by the pure state
\begin{equation}
	| \psi \rangle = \sum_{\alpha_1 = 0}^{1} \dotso \sum_{\alpha_n = 0}^{1} c_{\alpha_1, ..., \alpha_n} |\alpha_1, ..., \alpha_n \rangle,
\end{equation}
with $c_{\alpha_1, ..., \alpha_n} \in \mathbb{C}$.
For this state, the GME \cite{WG2003} of $|\psi\rangle$ is given by $E = 1 - \Lambda_{\text{max}}^2$, where $\Lambda_{\text{max}}^2$ is known as the \textit{entanglement eigenvalue}, and corresponds to the maximal fidelity $F(|\psi\rangle,|\phi\rangle)$ between the state $|\psi\rangle$ and the set $\{|\phi\rangle\}$ of pure separable state, that is,
\begin{equation}\label{lambda_max}
	\Lambda_{\text{max}}^2 = \max_{\{|\phi\rangle\}} F(|\psi\rangle,|\phi\rangle).
\end{equation}

\begin{figure}[t!]
        \centering
		\includegraphics[width=1\columnwidth]{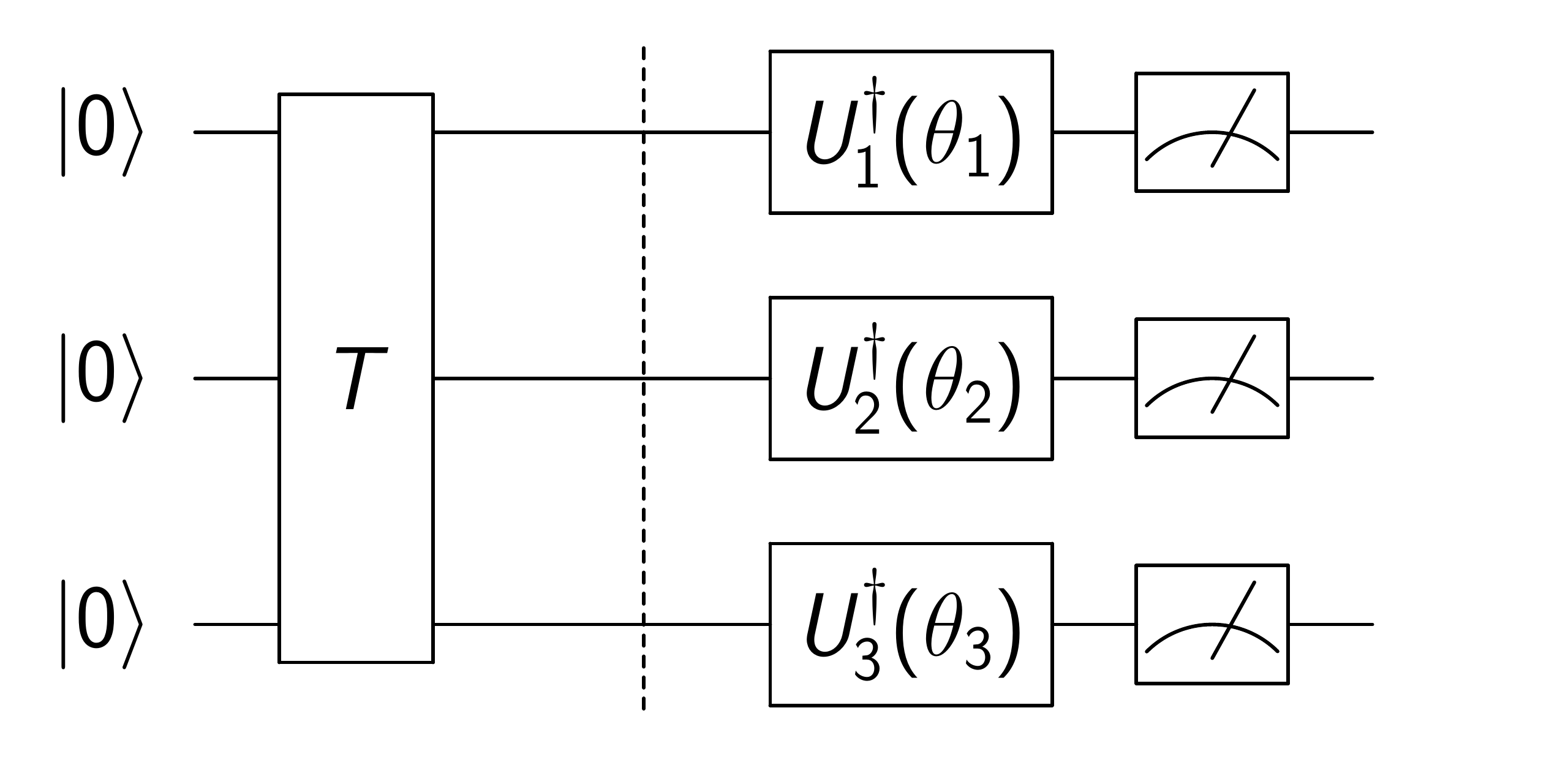}
		\caption{VDGE quantum circuit for an arbitrary 3-qubit state. The target state $\ket{\psi}$ is prepared applying a unitary gate $T$ onto the separable state $\ket{0}^{\otimes 3}$. The fidelity $|\braket{\phi(\boldsymbol\theta)}{\psi}|^2$ is obtained applying local unitary gates $U_i^{\dagger}(\boldsymbol{\theta}_i)$ and measuring all qubits in the computational basis.}
		\label{fig:qcircuit}
\end{figure}

We solve this optimization problem following a variational approach as \cite{F2014,URND2019}. The VDGE requires two basic quantum steps: (i) a variational ansatz for separable states, which in our case is constructed with $n$ single-qubit unitary transformations acting on the separable state $\ket{0}^{\otimes n}$, that is,
\begin{equation}
    \ket{\phi(\boldsymbol{\theta})} = U_1 (\boldsymbol{\theta}_1) \otimes \dots \otimes U_n (\boldsymbol{\theta}_n)\ket{0}^{\otimes n},
\end{equation}
where the vector $\boldsymbol{\theta}=(\boldsymbol{\theta}_1,\dots,\boldsymbol{\theta}_n)$ contains the parameters defining the action of the unitary transformations, and (ii) measuring the fidelity as a function of $\boldsymbol{\theta}$. Steps (i) and (ii) can be naturally implemented in a quantum computer. Here, each unitary $U_i(\boldsymbol{\theta}_i)$ is generated as a sequence of local quantum gates acting on the i-th qubit and the fidelity is obtained applying the operator $U^\dagger(\boldsymbol{\theta})=U_1^\dagger(\boldsymbol{\theta}_1) \otimes \dots \otimes U_n^\dagger (\boldsymbol{\theta}_n)$ onto the entangled state $|\psi\rangle$, followed by a projection onto the computational basis, as is shown in Fig.~\ref{fig:qcircuit}. The value of the fidelity $F(\boldsymbol{\theta})$ is estimated as
\begin{equation}
F(\boldsymbol{\theta}) = \qty|\braket{\phi(\boldsymbol{\theta})}{\psi}|^2=|\langle 0|^{\otimes n}U^
\dagger(\boldsymbol{\theta})|\psi\rangle|^2\approx \frac{n_0}{N},
\end{equation}
where $n_{0}$ is the number of counts obtained when projecting onto the state $\ket{0}^{\otimes n}$ and the ensemble size $N$ is the total number of copies of $\ket{\psi}$ used in the projective measurement.

Thereby, the calculation of the GME reduces to run an optimization loop in which we use a classical optimizer to maximize $F(\boldsymbol{\theta})$ by varying the ansatz's parameters $\boldsymbol{\theta}$. Once convergence is reached, we replace the optima $\boldsymbol{\hat\theta}$ in Eq.~\eqref{lambda_max} and calculate the GME as $\hat E = 1 - \Lambda_{\text{max}}^2 (\boldsymbol{\hat\theta})$.

Since the method uses experimental estimates of the fidelity in each step, it is affected by several sources of noise. These include errors due to state preparation and measurement (SPAM), decoherence, and stochastic fluctuations produced by finite sampling. This scenario restricts the choice of the optimization algorithm. In our case, we use CSPSA as classical optimizer. It has been proven that this method is robust to noise and it only requires two evaluations of the cost function per iteration, making it a suitable optimization algorithm for our problem. Given that CSPSA works in the field of complex variables, we set $\boldsymbol{\theta}$ to be a vector of $2n$ complex numbers, where each pair of parameters defines a single qubit state. This is a linear scaling in the number of parameters $O(n)$, which makes the method well suited for large numbers of qubits. CSPSA provides a sequence of estimates $\boldsymbol{\hat\theta}_{k}$ that converges to the minimizer $\boldsymbol{\hat\theta}$ of the fidelity. At a given iteration $k$, a new estimate $\boldsymbol{\hat\theta}_{k+1}$ is generated from the fidelity values $F(\boldsymbol{\hat\theta}_{k,\pm})$ at vectors $\boldsymbol{\hat\theta}_{k,\pm}$, which are generated from the previous estimate $\boldsymbol{\hat\theta}_{k}$. At the last iteration the fidelity $F(\boldsymbol{\hat\theta}_{k+1})$ is measured, which leads to the estimate $\hat E$ of the GME. 

A drawback present in the evaluation of the GME is its landscape, which may contain several local maxima. This can cause the optimization algorithm to be trapped in a local maximum. Also, the maximum may not be unique. For example, both states $|00\rangle$ and $|11\rangle$ maximize Eq.~\eqref{lambda_max} for the Bell state $(|00\rangle + |11\rangle)/\sqrt{2}$, giving the same value for the GME. A small modification in the parameters of this state gives us a landscape with local maxima in which the optimization algorithm can get trapped. To overcome this problem, we employ a multi-start strategy, where the algorithm is repeated several times. Since CSPSA is a stochastic optimization method, it approaches $\boldsymbol{\hat\theta}$ from different paths in search space, which leads to a set of estimates $\{\hat E_j\}$. The highest value in $\{\hat E_j\}$ is selected as the final estimate $\hat{E}_*$.

\begin{figure}[t]
        \centering
		\includegraphics[width=1\columnwidth]{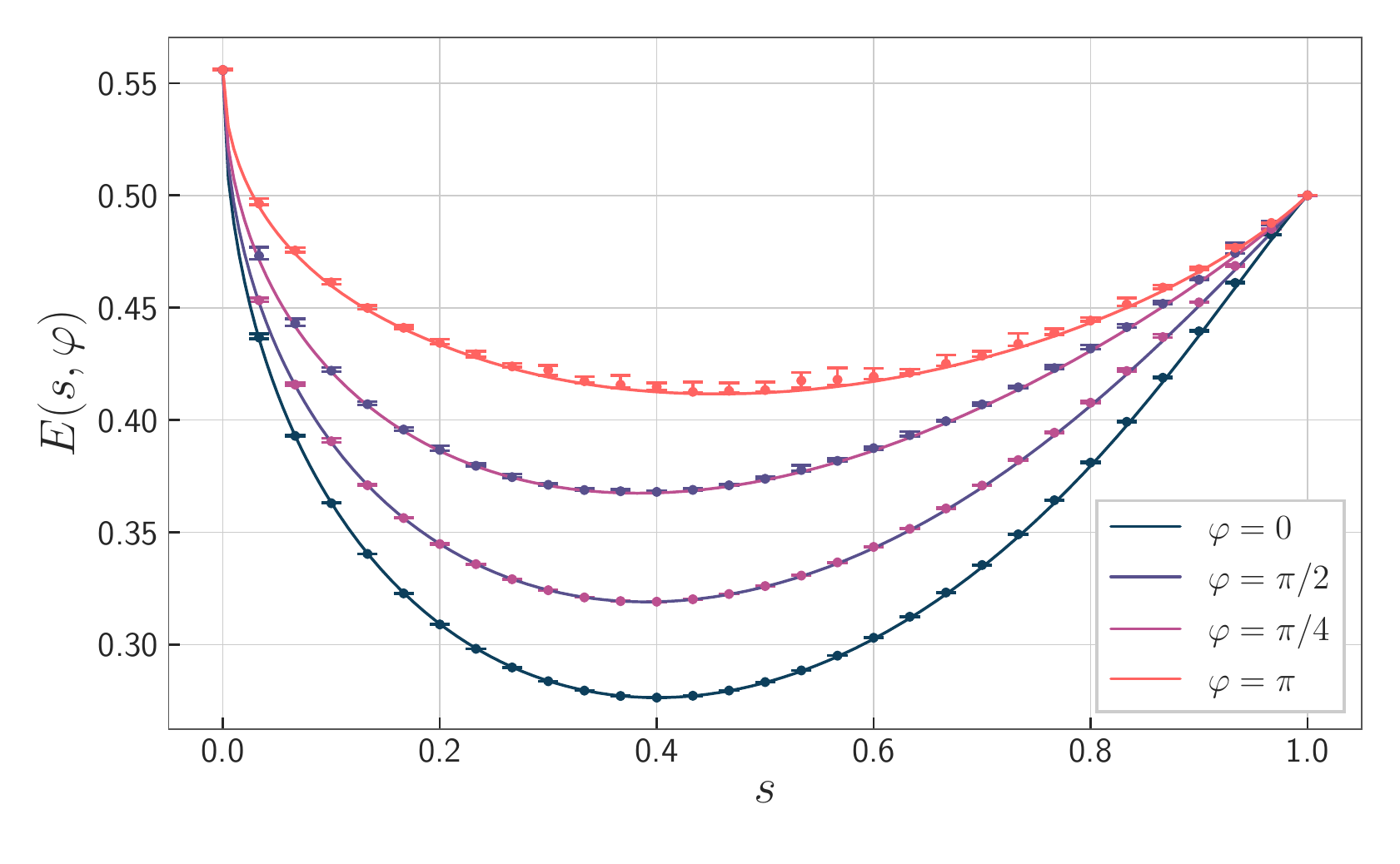}
		\caption{Geometric entanglement measure of $\ket{\mathrm{GW} \qty(s, \varphi)}$ as a function of $s$ for $\varphi=0,\pi/4,\pi/2,\pi$, from bottom to top. Solid lines correspond to the theoretical solution of the optimization problem using a classical optimization algorithm i.e. Basin-hopping. Dots represent the median value of the GME obtained using VDGE with $5$ repetitions and choosing the best result, while error bars correspond to the interquartile range.}
		\label{fig:ghzw}
\end{figure}

\section{Results and discussion}
\subsection{Numerical simulations}

\begin{figure*}[ht!]
  \begin{subfigure}[b]{1\columnwidth}
    \includegraphics[width=1\textwidth]{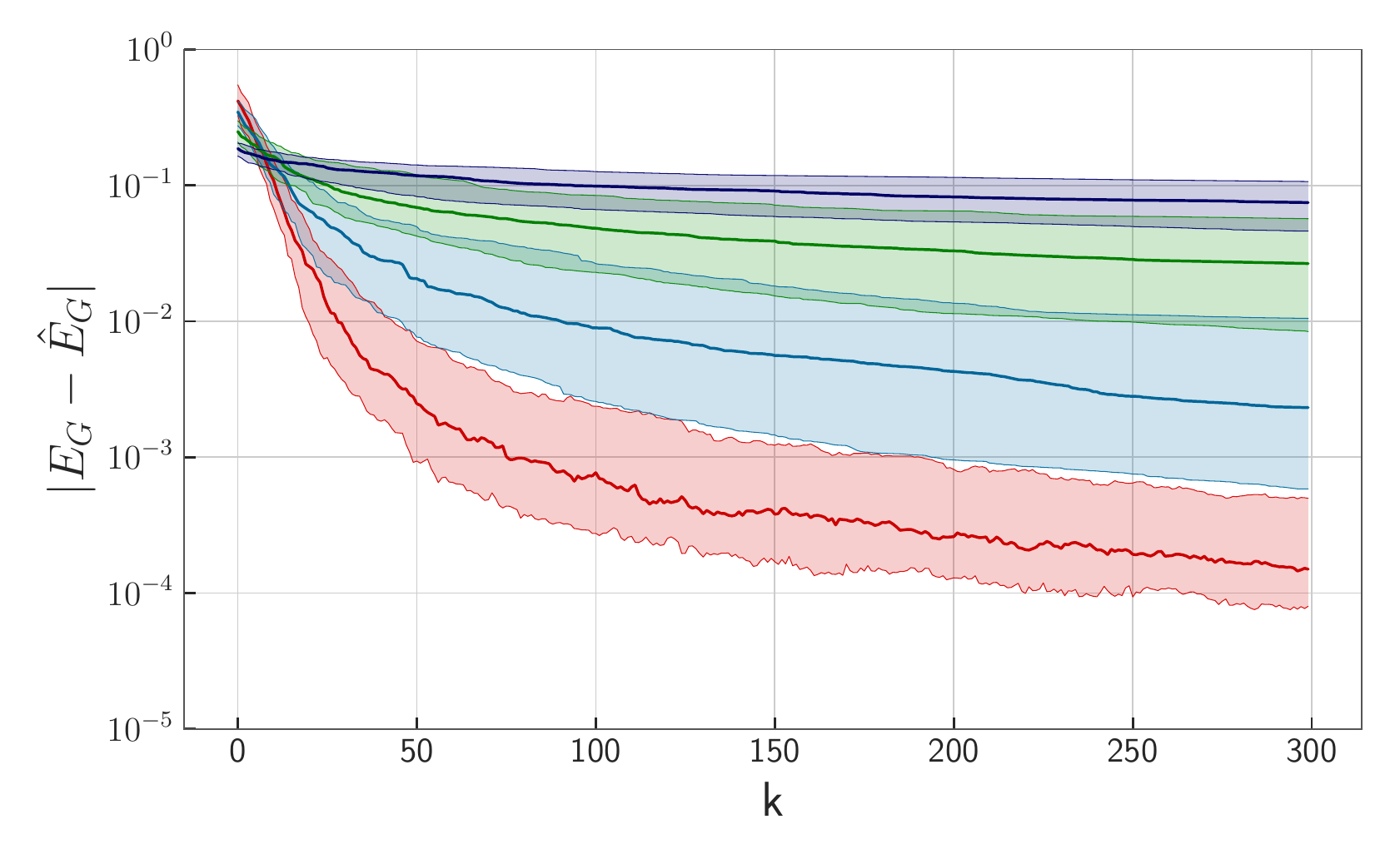}
    \caption{5 repetitions}
    \label{fig:medgeoGME and the variational GME 5}
  \end{subfigure}
  \begin{subfigure}[b]{1\columnwidth}
    \includegraphics[width=1\textwidth]{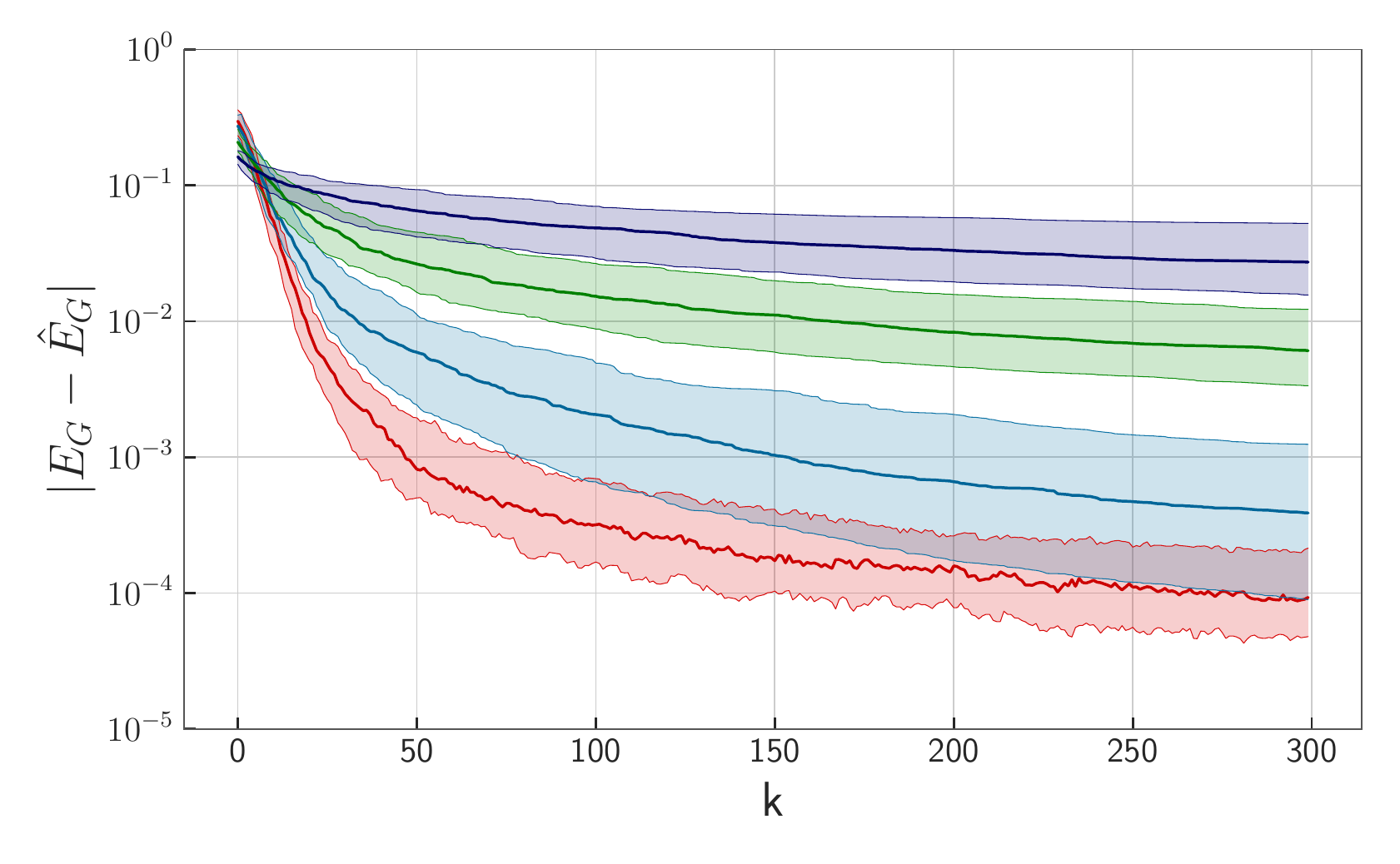}
    \caption{20 repetitions}
    \label{fig:medgeoGME and the variational GME 20}
  \end{subfigure}
\caption{Difference between the GME $E$ obtained with the Basin-hopping optimization method and the estimated GME $\hat{E}$ obtained with VDGE for 100 random pure states versus the number of iterations $k$. The curves show the results for $3$, $4$, $5$, and $6$ qubits from bottom to top. Solid lines denote the median difference, while shaded areas represent the corresponding interquartile range. The left figure uses a sample size of $5$ different initial states, while the right one uses $20$ initial states, showing a better performance.}
\label{fig:medgeoGME and the variational GME}
\end{figure*}

\begin{figure}[ht!]
\includegraphics[width=1\columnwidth]{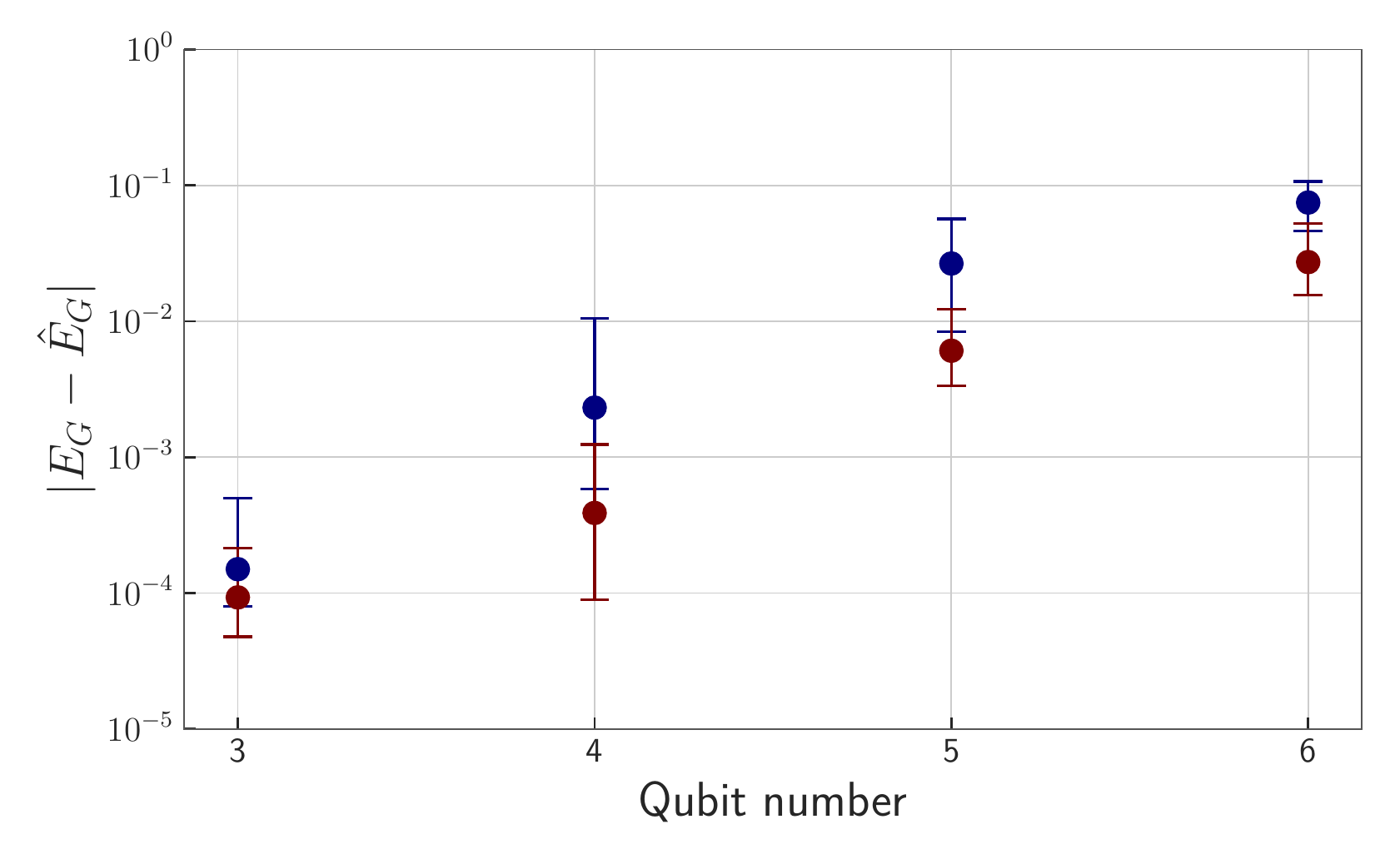}
\caption{Difference between the GME $E$ obtained with the Basin-hopping optimization and the estimated GME $\hat{E}$ obtained with VDGE versus the number of qubits, using a sample size of 5 initial states (blue) and 20 initial states (red). Dots correspond to the median difference of the last value obtained with VDGE and the theoretical value, while error bars represent the corresponding interquartile ranges.}
\label{fig:fid_vs_qubits}
\end{figure}

We first test the performance of our method with states of $n=3$ qubits. In particular, we simulate VDGE for a superposition of GHZ and W states given by \cite{WG2003}
\begin{equation}
    \ket{\mathrm{GW} \qty(s, \varphi)} = \sqrt{s}\ket{\mathrm{GHZ}} + e^{i \varphi} \sqrt{1-s}  \ket{ \mathrm{W} },
\end{equation}
where $s \in [0, 1]$, $\varphi$ is a relative phase, and
\begin{align}
\ket{\mathrm{GHZ}} &= \frac{1}{\sqrt{2}}( \ket{0}^{\otimes n} + \ket{1}^{\otimes n} ), \\
\ket{\mathrm{W}} &= \frac{1}{\sqrt{n}}( \ket{1\cdots0}+\cdots +\ket{0\cdots1}). 
\end{align}
For each of the values $\varphi=0,\pi/4,\pi/2,\pi$ we generate 31 equally spaced values of $s$, leading to a total of 124 $\ket{\mathrm{GW}\qty(s,\varphi)}$ states. The GME of each one of these states is calculated with VDGE using 100 repetitions, each one given by an initial condition chosen according to a Haar-uniform distribution. Each repetition consists of 150 iterations of CSPSA where the values of the fidelity are simulated employing a sample of $2^{13}$ shots, a number commonly available in open hardware like IBM Quantum. In order to estimate errors, we employ the bootstrapping method \cite{boots}. The generated data is used to calculate the median and interquartile range for each simulated state. These are summarized in Fig.~\ref{fig:ghzw}, where we show the value of the GME achieved by VDGE as a function of $s$ for the values  $\varphi=0,\pi/4,\pi/2,\pi$ from bottom to top. In this figure, points represent the median value of the GME and bars correspond to the interquartile range. Solid lines correspond to the theoretical solution of the optimization problem Eq.~\eqref{lambda_max} obtained using the Basin-hopping global optimization algorithm \cite{bashop}. As is apparent from this figure, VDGE generates median values that are almost identical to the theoretical solutions and interquartile ranges that are very narrow, with mean errors of $0.00011, 0.00091, 0.00022$ and $0.00124$ for the respective values of $\varphi=0,\pi/4,\pi/2,\pi$. Thereby, our simulations indicate that VDGE generates accurate values of the GME for all the simulated states.

We also ran numerical simulations to test our method with randomly generated states in various dimensions. Similar to the previous simulation, for $3$, $4$, $5$, and $6$ qubits we generate a set of $100$ random pure states according to a Haar-uniform distribution. For each state, the GME is selected as the maximum value obtained via VDGE over 5 and 20 repetitions using $2^{13}$ shots to simulate the measurement of the fidelity. We compare our results with the value obtained using  Basin-hopping. In particular, we calculate the median of $|\hat{E} - E|$ between the value $\hat{E}$ of the GME estimated via VDGE and the value $E$ of the GME obtained via the Basin-hopping algorithm. The results of these simulations are presented in Figs.~\ref{fig:medgeoGME and the variational GME 5} and \ref{fig:medgeoGME and the variational GME 20}. These show the median of $|\hat{E} - E|$ (solid lines) as a function of the number of iterations of CSPSA, using $5$ and $20$ repetitions, respectively, for $3$, $4$, $5$ and $6$ qubits from bottom to top. Shaded areas represent the interquartile range. As is evident from Figs.~\ref{fig:medgeoGME and the variational GME 5} and \ref{fig:medgeoGME and the variational GME 20}, the error in the estimation provided by VDGE decreases as the number of iterations of the optimization algorithm increases. Each curve exhibits a rapid decrease within the first tens iterations followed by approximately linear asymptotic behavior. This holds for all simulated qubit numbers. However, the quality of the estimation decreases as the number of qubits increases. This is depicted in Fig.~\ref{fig:fid_vs_qubits} that shows the error in estimating the GME as a function of the qubit number for $5$ and $20$ repetitions. Here, the result at iteration $k=150$ of the optimization process is compared to the theoretical value of the GME. This behavior is to be expected because the dimension increases exponentially with the number of qubits and the measurement of fidelity is simulated with a constant number $2^{13}$ of shots. The main difference between Figs.~\ref{fig:medgeoGME and the variational GME 5} and \ref{fig:medgeoGME and the variational GME 20} is that errors and dispersion are slightly less pronounced for $20$ repetitions, but in both cases, the algorithm converges successfully when tested with random pure states. 

\begin{figure}
    \centering
    \includegraphics[width=\linewidth]{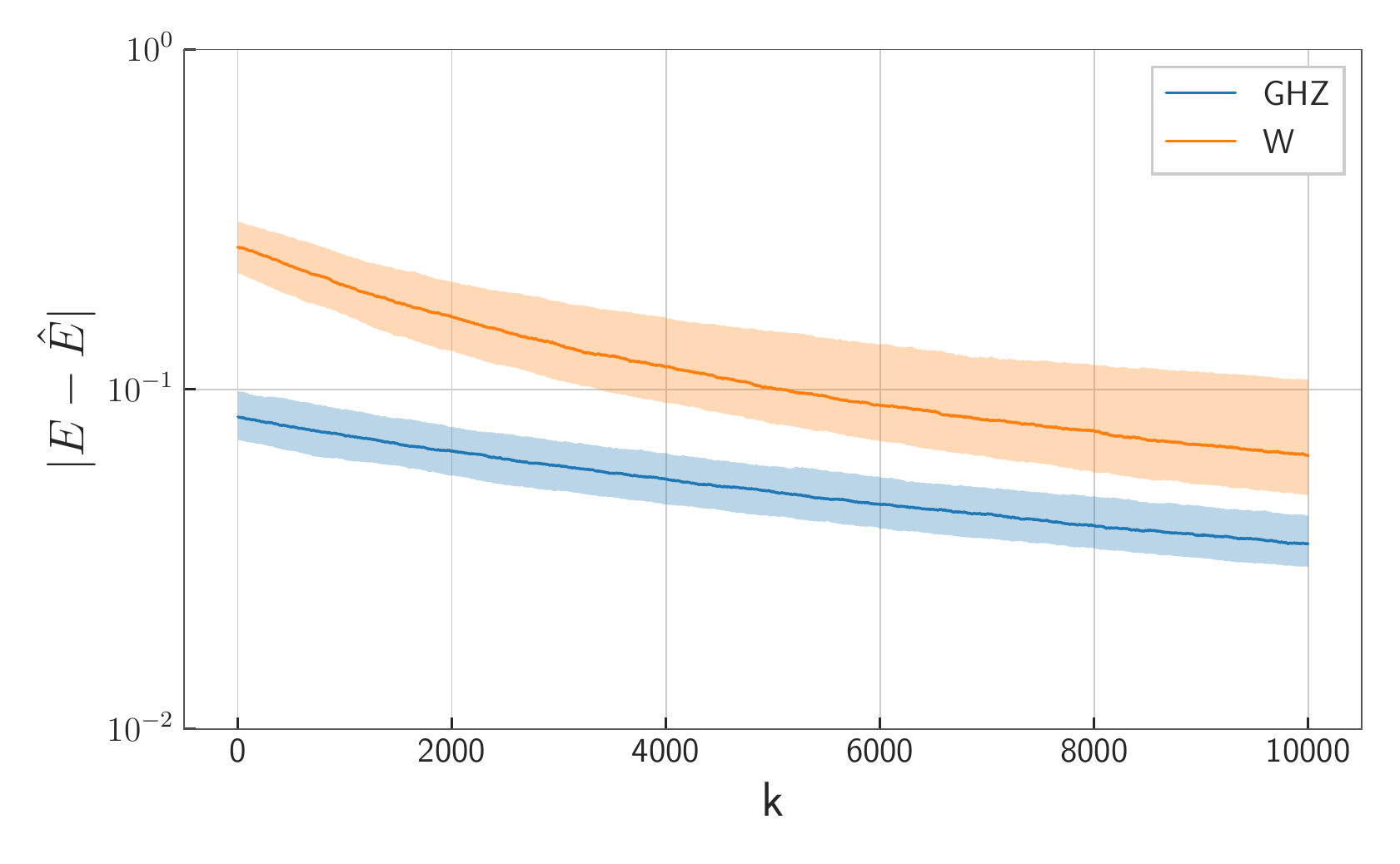}
    \caption{Median of the difference between the GME $E$ obtained with the Basin-hopping optimization method and the estimated GME $\hat{E}$ obtained with VDGE in a simulation based on matrix product states techniques for $10^3$ perturbed W state (upper orange curve) and GHZ state (lower blue curve) for 25 qubits, as a function of the number of iterations $k$. Shaded areas represent the interquartile range.}
    \label{fig:MPS}
\end{figure}

The previous simulations are restricted to small number of qubits. In order to extend our simulations to higher qubit numbers, we employ tensor network algorithms \cite{ORUS2014117}. These have proven to be a useful tool for performing numerical simulations in many-body quantum systems \cite{Klumper_1991,Fannes1992,PhysRevLett.69.2863,Kl_mper_1993,PhysRevB.48.10345,PhysRevLett.93.227205,GarcaRipoll2021}. In particular, we carry out the simulations over matrix product states (MPS), that is, states that can be written as
\begin{align}
    |\psi\rangle = \sum_{j_a,i_b} A^{(1)}_{j_1i_1}A^{(2)}_{i_1j_2i_2}\dots A^{(n)}_{i_{n-1}j_n}\ket{j_1\dots j_n},
\end{align}
where $A^{(j)}$ are rank-$3$ tensors, with $j=1,\dots,n$. Algorithms have been proposed to efficiently compute the GME of MPS \cite{Hu_2011,Teng2017}. We focus our study on states in the neighborhood of GHZ and W states, which have an efficient MPS representation. To generate the probe MPSs, we perturb the tensors $A^{(j)}$ of GHZ and W states with random matrices generated by a Gaussian distribution of null mean and variance $\lambda$, and then normalize the MPS. Figure~\ref{fig:MPS} shows the median of $|\hat{E} - E|$ on a set of $10^3$ perturbed GHZ and W states of $n=25$ qubits, with $\lambda=0.1$. VDGE is executed with $10^4$ iterations of CSPSA simulating the fidelity with a sample of $2^{13}$ shots. We use as initial conditions the optimums of GHZ and W states without perturbation. We do this because the dimension of the space is too large, and therefore an unattainable number of shots is required to converge when a random initial condition is used. Figure~\ref{fig:MPS} shows that with only $2\times10^4$ fidelity evaluations, we can reduce the average error in estimating the GME by about half of an order of magnitude. This number of evaluations is a thousand times smaller than the dimension of the system $d=2^{25}\approx3\times 10^7$.

\subsection{Experimental results}

We performed an experimental demonstration of VDGE using IBM Quantum systems \texttt{ibmq\_lima} and \texttt{ibmq\_bogota}. In particular, we measure the geometric entanglement of a $n$-qubit GHZ state with $n=3,4,5$. VDGE is repeated 10 times  for each state
considering 150 iterations and $2^{13}$ shots for measuring the fidelity at each iteration.

\begin{figure}[ht!]
        \centering
        \begin{subfigure}[b]{1\columnwidth}
		\includegraphics[width=1.0\textwidth]{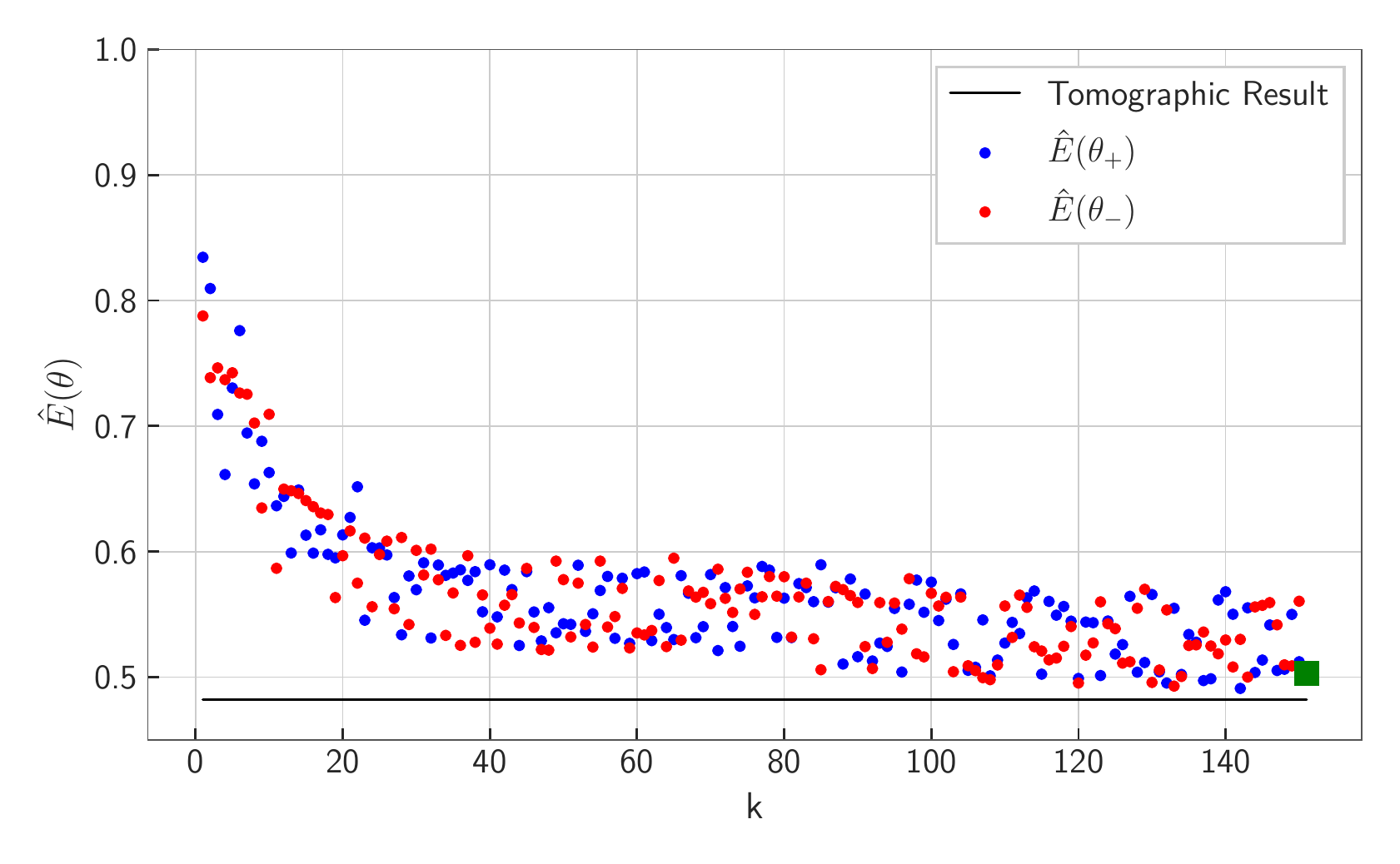}
		\end{subfigure}
		\begin{subfigure}[b]{1\columnwidth}
		\includegraphics[width=1.0\textwidth]{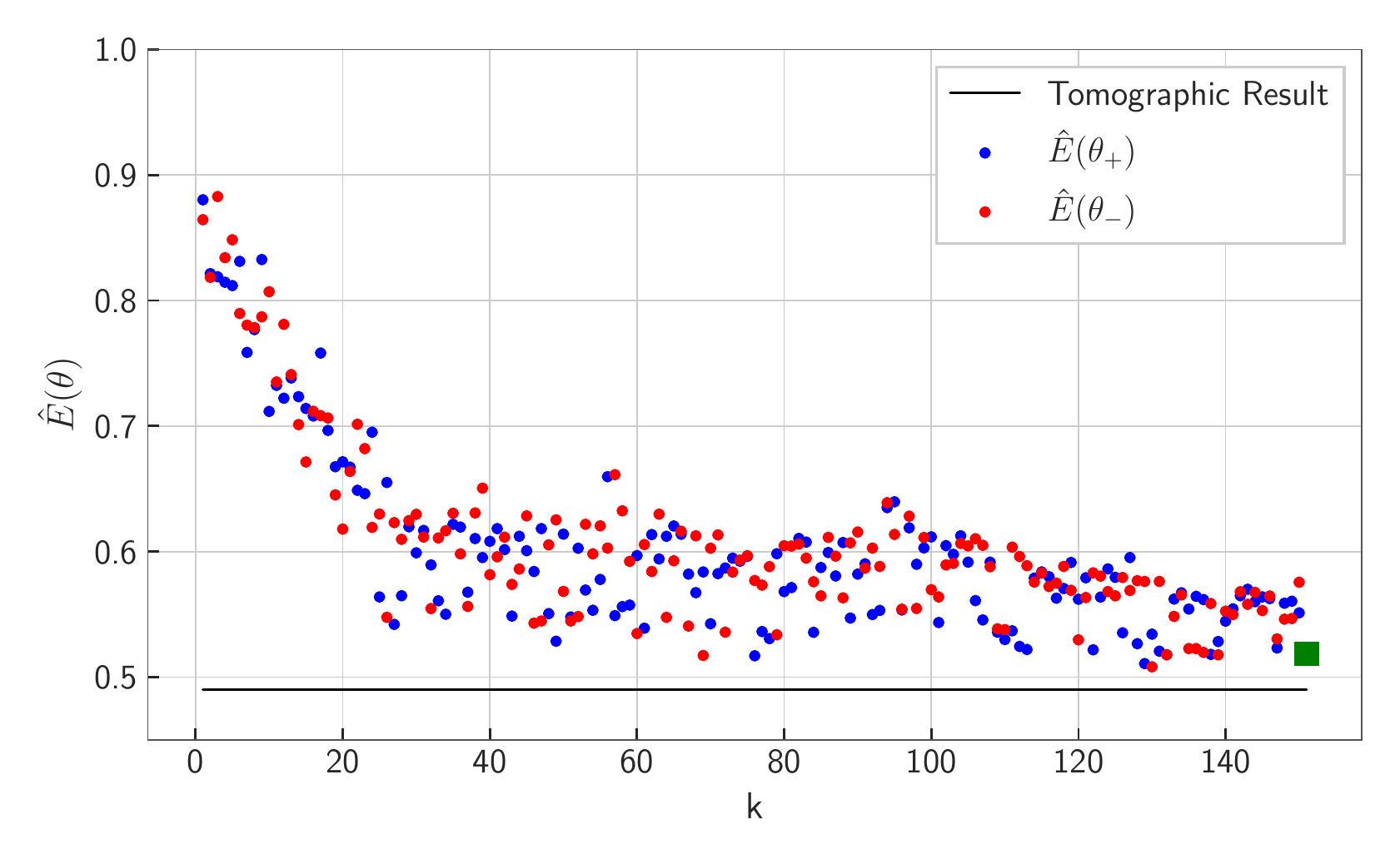}
		\end{subfigure}
		\begin{subfigure}[b]{1\columnwidth}
		\includegraphics[width=1.0\textwidth]{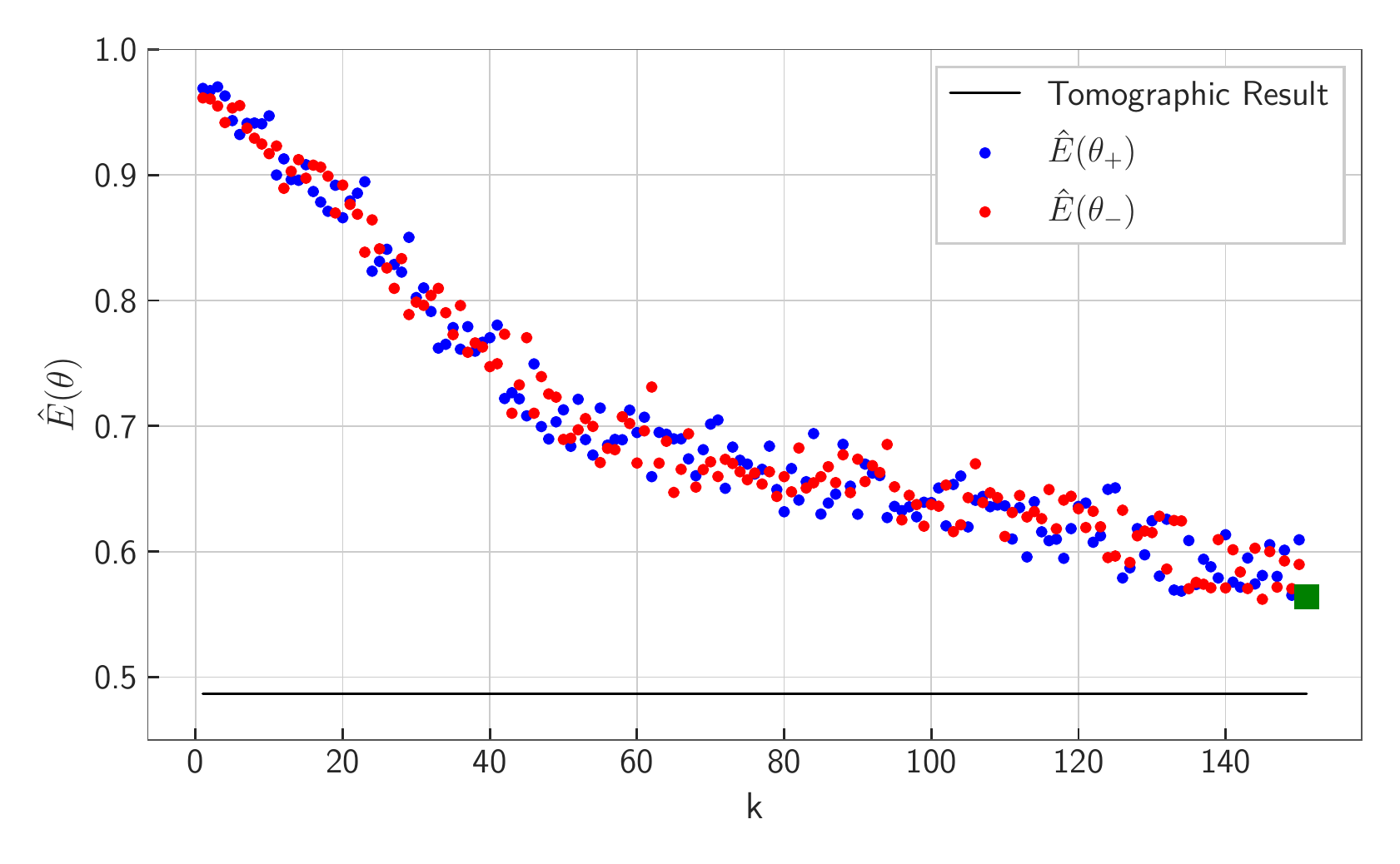}
		\end{subfigure}
	\caption{Values of $\hat{E}(\boldsymbol{\theta}_{k,+})$ (blue dots) and $\hat{E}(\boldsymbol{\theta}_{k,-})$ (red dots) as functions of the number of iterations $k$ for 3, 4, and 5 qubits, from top to bottom. The final estimate $\hat{E}_*$ of the GME at $k=150$ obtained via VDGE is indicated with a solid green square. Solid black lines indicate the value of the GME obtained by reconstructing the generated states with standard quantum tomography and then solving the optimization problem via Basin-hopping.}
	\label{fig:exp}
\end{figure}

The results of our experiment are illustrated in Fig.~\ref{fig:exp} for 3, 4, and 5 qubits, from top to bottom. In each inset we depict the values $\hat{E}(\boldsymbol{\theta}_{k,+})$ (blue dots) and $\hat{E}(\boldsymbol{\theta}_{k,-})$ (red dots) as functions of the number of iterations $k$. The value of the final estimate is indicated as a green square at the last iteration.
The values of $\hat{E}(\boldsymbol{\theta}_{k,\pm})$ exhibit a rapid decrease within the first ten iterations followed by approximately linear asymptotic behavior. In addition, the values of $\hat{E}(\boldsymbol{\theta}_{k,\pm})$ become very close as the number of iterations increases, which shows the convergence of the algorithm toward the maximum. The final estimates $\hat{E}_*$ are $0.5029$, $0.5184$, and $0.5640$ for $n=3,4,5$, respectively. These figures can first be compared with the value $E=0.5$, which is the theoretical value of the GME for a GHZ state, independently of the number of qubits. VDGE provides a value of the GME which is close to the theoretical value with relative errors of $0.0058$, $0.0368$, and $0.128$, respectively. The relative error increases as the number of qubits increases, which can be explained by the increase in the dimension of the search space. However, as the algorithm enters and stay into the linear regime it is possible to reduce the relative error by increasing the number of iterations. The GHZ state is generated by concatenating several Control-Not gates. This gate has an error larger than local gates and its impact on the generated state and the GME value can be significant. Therefore, we reconstruct the generated state via standard quantum tomography and forced purity and calculate its GME via Basin-hopping. This leads to the GME values $0.48214$, $0.48996$, and $0.4868$ for the generated states for $n=3,4,5$ qubits, respectively, which in Fig.~\ref{fig:exp} are indicated with continuous black lines. With respect to these values, VDGE has relative errors of $0.0431$, $0.0581$, and $0.1586$, correspondingly.

Overall, VDGE exhibits convergence and provides in our experiment errors in the order of $10^{-2}$. This is larger than what was observed in the numerical simulations in Fig. ~\ref{fig:medgeoGME and the variational GME}. This can be explained by the single-qubit gate error and readout assignment error, which are in the order of $10^{-4}$ and $10^{-2}$, respectively. Since VDGE relies solely on single-qubit gates, its implementation in current NISQ computers is feasible.

\section{Conclusions}

Multipartite entanglement plays a key role in quantum information and quantum communication. However, the theoretical characterization of entanglement, as well as its experimental determination and certification, appear to be difficult problems. This is mainly due to the fact that most entanglement measures are not directly related to physically measurable quantities and are difficult to calculate, especially for higher-dimensional quantum systems. 

Here, we study the problem of measuring the entanglement of multipartite pure states generated in NISQ computers. This class of devices is characterized by noisy quantum gates that severely limit their usefulness. In this context, a major research subject is the development of algorithms that can work under such disadvantageous conditions. 

We use the geometric measure of entanglement, which characterizes the entanglement of a $n$-qubit pure state $|\psi\rangle$ as the distance to the nearest separable pure state. This entanglement measure can be calculated by maximizing the fidelity of $|\psi\rangle$ in the set of separable states. The fidelity can be experimentally obtained in a NISQ computer by projecting $|\psi\rangle$ onto a set of $n$ local bases. The complex simultaneous perturbation stochastic approximation algorithm is then used to solve the optimization problem in a classical computer. Thus, we provide a variational determination of the geometric entanglement measure. In addition to its purity, the method does not require a priori information about the state $|\psi\rangle$. The GME is well suited for being implemented in NISQ computers. This is because the measurements necessary to evaluate the GME are local and require a single unitary transformation that acts on each qubit. Thus, VDGE has circuit depth 1, which helps to decrease error accumulation.

Numerical simulations indicate that our method reproduces known results on the geometric measure of entanglement of superpositions of 3-qubit GHZ and W states. Simulations with randomly chosen states of 3, 4, 5, and 6 qubits show that our method can provide accurate values of the geometric entanglement, which is controlled by the number of iterations and the size of the ensemble used to measure the fidelity. We also extended our simulations up to 25 qubits using matrix product state techniques, where we also observe convergence towards the optimum with a reasonable number of fidelity evaluations despite the large size of the dimension.

We also carry out an experimental demonstration of the variational determination of multipartite entanglement using IBM Quantum \texttt{ibmq\_lima} and \texttt{ibmq\_bogota} devices. In particular, we calculate the geometric entanglement measure of a GHZ state for 3, 4, and 5 qubits and obtain relative errors in the order of $10^{-2}$ for 3 and 4 qubits and in the order of $10^{-1}$ for 5 qubits. Our demonstration uses 150 iterations and 10 repetitions. The simulations indicate that an increase in these figures leads to a reduction of relative error. Another important factor is the size of the ensemble used to measure the fidelity, which is currently limited to $2^{13}$ shots. A consequence of this is that the accuracy in the estimation of the fidelity decreases as the number of qubits increases, which in turn reduces the convergence rate of the VDGE. 

Our results find direct application in the entanglement quantification of high-dimensional pure entangled states. Recently, two quantum-hardware platforms have been used to generate high-dimensional entanglement. The IBM Quantum \texttt{ibmq\_montreal} device has reportedly generated \cite{ghz27} a 27-qubit GHZ state and detected its genuine multipartite entanglement. On a similar platform, the IBM Quantum \texttt{ibmq\_rome} device, the generation of a three-qutrit GHZ state has been carried out for the first time \cite{qutrits} and its genuine multipartite entanglement has also been demonstrated. Also recently, a universal qudit quantum processor using trapped ions \cite{trapions} has been experimentally demonstrated. Here, individually addressable qudits up to dimension 7 are created and accurately controlled. These quantum devices also allow the implementation of our approach, which makes it possible to characterize the entanglement of arbitrary states of a large number of qubits and even qudits. In particular, high-accuracy local gates are implemented in \cite{trapions}, which is the basic resource for the variational determination of the geometric measure of entanglement on any dimension.

Our proposal also provides advantages over classical numerical techniques to study the quantum entanglement of multipartite systems. The most promising are algorithms based on tensor networks \cite{Wei2005global,Orus2008universal,Shi_2010,Roy2019tensor}, which allowed unprecedented access to the physical properties of many-body systems. However, they only work for bounded families of entangled states or for particular quantum systems. VDGE is able to estimate the entanglement of any state, particularly those with long-range correlations that cannot be efficiently represented by tensor networks.

The VDGE can be extended to estimate the genuine entanglement of a multiqubit state \cite{Kraft2018char,Guo2018exp, Friis2018}. This requires considering parameterized states of the form $ |\psi_{A} \rangle \otimes |\psi_{B}\rangle $ for all $A, B \subseteq \{1, ..., N\}$, $A \cap B = \emptyset$, $A, B  \neq \emptyset$. Using these states, we could find the maximum overlap with the set of bi-separable states. If the overlap is different from one, we can ensure that the state is genuinely entangled. Since there are $O(2^{N})$ ways in which we can distribute $N$ distinguishable qubits into two non-empty sets, this procedure would greatly increase the complexity of the algorithm. However, knowledge about the physical geometry of the qubits could decrease this complexity. Another possible extension of our method is the estimation of entanglement of mixed state. Nevertheless, this has proven to be a difficult problem because generalizations of the GME to mixed states are based on the convex roof extension \cite{Bennett1996mixed,WG2003,Toth2015eval,Horodecki2009quantum} or on finding the closest separable density matrix \cite{re4, re1, Spehner2017, Girardin2021build}. Both approaches involve optimization problems with an exponential number of parameters in the number of qubits, making it challenging to solve them even in the case of small systems. Extensions to genuine entanglement and mixed states require extra considerations which will be dealt with in future works.


\begin{acknowledgments}

This work was supported by ANID -- Millennium Science Initiative Program -- ICN17$_-$012. AD was supported by FONDECYT Grant 1180558. LP was supported by ANID-PFCHA/DOCTORADO-BECAS-CHILE/2019-72200275 and by the Spanish project PGC2018-094792-B-I00 (MCIU/AEI/FEDER, UE). LZ was supported by ANID-PFCHA/DOCTORADO-NACIONAL/2018-21181021. J. C.-V. was supported by ANID-PCHA/DoctoradoNacional/2018-21181692. We acknowledge the use of IBM Quantum services for this work. The views expressed are those of the authors, and do not reflect the official policy or position of IBM or the IBM Quantum team.

\end{acknowledgments}

\bibliography{bibtex.bib}

\end{document}